Dec 2013

# Transiting Exoplanet Simulations with the James Webb Space Telescope

**Natasha Batalha[1], Jason Kalirai[1], Jonathan Lunine[2], Mark Clampin[3], Don Lindler[3]**
[1]*Space Telescope Science Institute, 3700 San Martin Dr. Baltimore, MD 21218*
[2]*Department of Astronomy, Cornell University, 610 Space Science Building, Ithaca, NY 14853*
[3]*Goddard Space Flight Center, 8800 Greenbelt Rd, Greenbelt, MD*, 20771

## Introduction

The James Webb Space Telescope (JWST) was ranked as the top science priority in the 2000 Astronomy & Astrophysics Decadal Survey. The telescope has a 6.5 meter primary aperture with sensitivity from 0.6 to 28 microns and has four science instruments with sensitivity between 0.6 and 28 microns. Offering orders of magnitude gains in sensitivity and imaging and spectroscopic resolution, JWST is expected to achieve breakthroughs in many astrophysical disciplines.

Since the year 2000, our basic understanding of the demographics of planets around nearby stars has been completely transformed. Today, new exoplanet candidates are being discovered almost every day. The most significant result comes from the *Kepler* mission, which has now accumulated over 2000 planet candidates (Batalha et al. 2013). The Kepler data demonstrate an abundant population of "super-Earths" orbiting low mass, nearby stars. Of particular interest are systems such as Gliese 667C (Anglada-Escudé et al. 2012), Gliese 581 (Vogt et al. 2010), Tau Ceti (Tuomi et al. 2013), HD 85512 (Pepe et al. 2011) and HD 40307 (Mayor et al. 2009), all of which are closer than 15 pc and believed to harbor sub-Neptunian habitable zone planets.

The discovery of exoplanets around nearby stars opens the door to a new scientific front: exoplanet characterization. Both Hubble and Spitzer have made initial contributions in this area by observing atmospheres of planets with large hydrogen-helium envelopes, Jupiter-analogs, or planets with hot surface temperatures (Crouzet et al. 2012, Sing et al. 2011, Deming et al. 2007). Unfortunately, these telescopes lack the power and resolution to extend this exciting work to super-Earth planets with $M = 1\text{-}10\ M_{\oplus}$.

In this white paper, we assess the potential for JWST to characterize the atmospheres of super-Earth exoplanets, by simulating a range of transiting spectra with different masses and temperatures. Our results are based on a JWST simulator tuned to the expected performance of the workhorse spectroscopic instrument NIRSpec, and is based on the latest exoplanet transit models by Howe & Burrows (2012). This study is especially timely since the observing modes for the science instruments on JWST are finalized (Clampin 2010) and because NASA has selected the TESS mission as an upcoming Explorer. TESS is expected to identify more than 1000 transiting exoplanet candidates, including a sample of about 100 nearby (<50 pc) super-Earths (Ricker et al. 2010).

In the following section 1, we provide a brief explanation of the JWST simulator. In section 2 we present the atmospheric models we used as input to the simulator. Section 3 provides the method used for determining the S/N of the simulated observations, and Section 4 summarizes the results. Section 5 provides an outline for future work that can be done with the simulator.

## 1. JWST Simulator

We use a JWST simulator that was developed by Mark Clampin and Don Lindler at GSFC. The JWST Simulator is comprised of several IDL routines, and takes input planetary transit depths and converts them to an output transit spectrum as would be observed by JWST. The simulator accepts a set of input parameters that include: the number of observed transits, duration of each transit, jitter and drift specifications, as well as a primary in and out of transit spectra. Line of sight requirements were under revision in the timeframe of this work and might need to be revisited for future studies. The out-of-transit model is simply a stellar spectrum. The in-transit spectrum is the same stellar spectrum convolved with a planetary absorption spectrum. The simulation begins by converting the input models from wavelength vs. flux (Å vs ergs/cm^2/sec/Å) to wavelength vs. count rates in 1/10 pixel bins in the dispersion direction using combined efficiencies, telescope effective area and wavelength dispersion. These can then be mapped onto an image sampled at 1/10 of a pixel by convolving with the Point Spread Function and multiplying by the Pixel Response Function(PRF). The PRF gives insights into how the pixels respond to starlight during a nominal observation and is based on real measurements of the FGS detectors. Several sources of error are then added to the image, including zodiacal and stray light, flat field errors, Poisson and read-noise. This procedure is repeated for each observation.

The NIRSpec simulator is split up into three bands: 1 – 1.9μm, 1.5 – 3μm, 2.75 – 5μm. The process of simulating a full spectrum requires large computing times. Each transit takes ~ 1 hour to simulate on a 3.4 GHz single processor with four cores. To extract the spectra, an average background is computed using the region outside of the extraction region. An optimal extraction procedure could improve the resultant spectrum, however, we do not consider optimal extraction techniques in this work.

## 2. Models

To produce the primary in and out of transit models, we use new theoretical transit spectra of super-Earths by Howe & Burrows (2012), and theoretical stellar spectra by Castelli and Kurucz (2004). Both sets of models are publically available. Howe & Burrows generic transit models are computed for a silicate-iron planet with the physical parameters of the GJ1214 system outlined in Table 1 and 2, except as specified. All of the simulations are computed for a planet orbiting an M-dwarf at 0.2 M⊙. They compute pure atmosphere models with compositions of water, carbon dioxide and methane as well as hydrogen rich atmosphere models for which solar abundances are multiplied by a single factor (which can be less than or greater than unity).

**Table 1**

Properties of Transiting Super-Earth GJ1214b (Charbonneau et al. 2009)

| Radius (R⊕) | Mass (M⊕) | *a* (AU) | Period (d) | *i* (deg) | *e* |
|---|---|---|---|---|---|
| 2.678±0.13 | 6.55±0.98 | 0.0143 | 1.5803925 | $88.62^{+0.36}_{-0.28}$ | <0.27 |

**Table 2**

Properties of Host Star GJ 1214

| $m_V$ | Radius (R⊙) | Mass (M⊙) | $T_{eff}$ (K) | Luminosity (L⊙) | [Fe/H] |
|---|---|---|---|---|---|
| 14.67 | 0.2110 | 0.157 | 3026 | 0.00328 | +0.39[2] |

We use three categories of super-Earths adapted from Miller-Ricci et al. (2009). The first category includes super-Earths with large surface gravities that are able to retain massive hydrogen atmospheres. We refer to this as the H-rich case. On the other end of spectrum we have a category of super-Earths that bear close resemblance to Earth, with atmospheres depleted in hydrogen and composed of mostly heavier molecules such as $CO_2$ and $H_2O$. We refer to this as the hydrogen poor case. A third case would be the super-Earths that have lost moderate levels of their atmospheric hydrogen due to either incomplete escape of hydrogen and/or outgassing of a significant secondary $H_2$ atmosphere (Miller-Ricci et al. 2009). Howe & Burrows do not have a model that represents this intermediate case but such objects are modeled in the literature. Therefore, the intermediate region is left for future work.

For a H-rich atmosphere, we use a 3x solar metallicity model which has a mean molecular weight of ~2.3 g/mole and, therefore, a relatively large scale height (Earth's MMW is 28.97 g/mole). For an atmosphere depleted in hydrogen and composed of mainly heavy elements, we use a pure water atmosphere with a mean molecular weight of 18 and therefore, a relatively small scale height. Refraction, which limits the altitude below which primary transits provide spectral information (García-Muñoz et al., 2012), can be neglected here as all transiting planets are assumed to be close to their parent stars, minimizing the effect.

## 3. Parameterizing the Capabilities of NIRSpec

For each atmosphere, we ran several simulations varying planet mass, planet temperature and distance from Earth separately. These parameters are outlined in Table 3. This allowed us to quantify the effect of each parameter on the resulting output spectrum, characterized through its signal-to-noise ratio (SNR). Then we interpolated over the full parameter space, which enabled us to estimate how well NIRSpec will be able to resolve whether a super-Earth has a hydrogen-rich or a hydrogen-poor atmosphere. It also helped us determine lower limits to how small, cool, or distant an object can be observed with NIRSpec.

**Table 3**

Howe & Burrows Input Parameter Choices and Distances

| Atmosphere | Temp (K) | Mass (M⊕) [Radius (R⊕)]* | Distance (pc) |
|---|---|---|---|
| 3xSolar | 400 700 1000 | 1[1] 2[1.2] 4[1.6] 7[1.9] 10[2.2] | 6 15 30 |
| $H_2O$ | 400 700 1000 | 1[1] 2[1.2] 4[1.6] 7[1.9] 10[2.2] | 6 15 30 |

*Howe & Burrows (2013) normalize their models to radius. Therefore, we assume an Earth density to calculate radius ($M^{\frac{1}{3}}$).

For the H-rich atmosphere looked at a $H_2O/CH_4$ absorption features centered at 1.4 μm, $H_2O$ at 1.8 μm and 2.7 μm, and $CH_4$ at 4.6 μm. For the H-poor atmosphere we looked at $H_2O$ absorption features at 1.4 μm, 1.8 μm and 2.7 μm. Since we aim to calculate the SNR for many models, we focus initially on the first band (1 – 1.9μm) of the spectra for the H-poor model and the second band (1.5 – 3μm) of the spectra for the H-rich model. This reduces the computing time for each simulation by one third.

To calculate the SNR of the spectra we fit the continuum using a standard first order polynomial scheme to remove the underlying slope in the specified wavelength region. Then we sum the flux contribution at each wavelength to get total signal and divide by the standard deviation of the noise times the square root of the number of points (RMS noise).

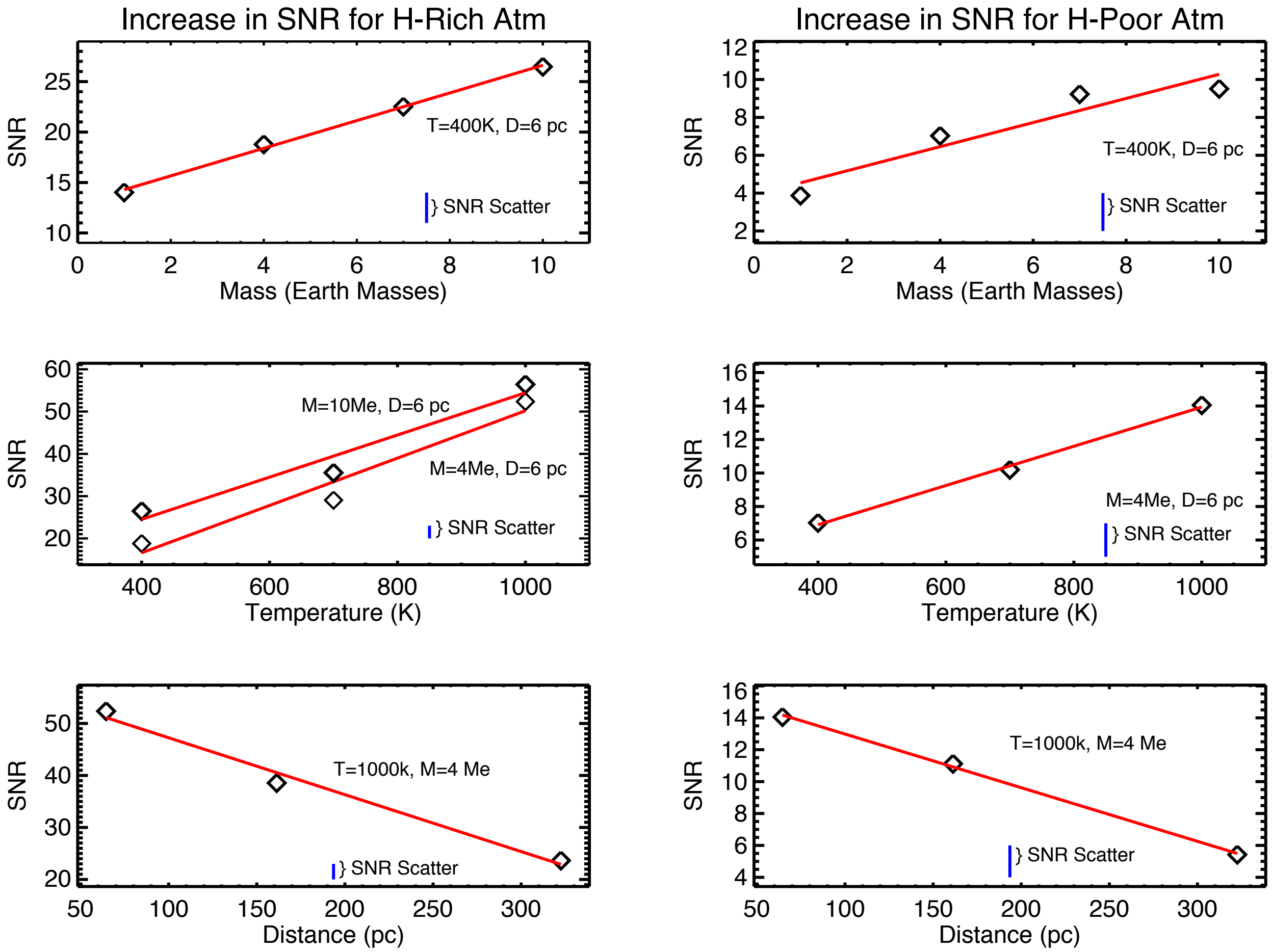

**Figure 1. SNR vs Planet Parameters**- SNR is plotted versus various planet parameters including planet mass, planet temperature and distance from Earth. Simulations are for NIRSpec at R~300, 25 transits, for different planets orbiting an M-dwarf. SNR calculations are described in Section 3. Diamonds represent calculated values and the red line is a first order polynomial fit through the points. Shown in the bottom right hand corner of each panel is the variation in SNR across the entire wavelength region (1 – 5 μm). The middle left hand plot shows two lines corresponding different masses to demonstrate that the slope of the temperature relationship is not strongly affected.

The sensitivity of NIRSpec is wavelength dependent, and, therefore, the SNR differs between the different bands defined above. We compute the SNR for each band separately. SNR for the entire spectrum is roughly √3 times the SNR for a single band since more features are included. **Figure 1** shows the simulated spectra as a function of increasing mass, temperature and distance with NIRSpec at R~300 at a single band. The figure shows how the SNR varies over the wavelength region (1 – 5μm). Each simulation was done using 25 transit observations.

Simulations were done for super-Earths in 1.5-day orbits, which places little constraint on the number of transit observations based on JWST's five-year mission lifetime. According to Kopparapu et al. 2013, super-Earths at the inner edge of the habitable zone of a 0.2 M⊙ M-dwarf will be in ~20-day orbits and those at the outer edge will be in ~45 day orbits. Therefore, twenty-five transits will even be obtainable for super-Earths in the habitable zones of their M-dwarf parent stars. Furthermore, based on JWST's five-year mission, a super-Earth in the outer edge of the habitable zone of a .2 M⊙ M-dwarf could be observed with up to 40 transits.

**4. Simulation/SNR Results**

Returning to **Figure 1**, the most striking difference lies between the SNRs of the H-poor and H-rich simulations (right side vs. left side – note the difference in scale). The highest SNR of the H-poor simulations (~14 for a 10 M⊕, 1000 K planet) is no greater than the lowest SNR of the H-rich simulations (~14 for a 1 R⊕, 400 K planet). This is an effect of the atmosphere's scale height, which is inversely proportional to the mean molecular weight of the gas. Planets with hydrogen-rich atmospheres have larger scale heights because of the small molecular weight. Such atmospheres are normally associated with gas giant planets that have no visible solid surface. Were we to assume an Earth with a hydrogen-rich atmosphere (that would, in fact, not be stable for long), the resulting S/N would be lower; likewise for an H-rich atmosphere that is cloudy.

**Figures 2** and **3** illustrate simulated spectra for a range of temperatures and masses based on the SNR calculations above. **Figure 2** shows NIRSpec observations for H-rich atmosphere models at 1 – 5μm assuming 25 transits as a function of temperature and mass. **Figure 3** displays the same information as **Figure 2** but for H-poor atmosphere models. Each of these plots is organized in a grid where each panel shows the output JWST simulated spectrum at a given planetary temperature (x-axis) and mass (y-axis). The red curve is the Howe & Burrows model as published; the black curve is the simulated spectrum at a distance of 4.5 pc. We also display a spectrum in green, which represents the furthest distance NIRSpec could observe the object and still maintain a SNR of at least ~15 (our threshold for detection; shown by the green arrow).

Looking at the case of the 10 M⊕, 1000 K, H-rich planet in the upper right hand panel of **Figure 2**, there is clear detection of $H_2O$ and $CH_4$ at all the wavelengths mentioned above at 4.5 pc (black curve). At 50 pc (green curve) the $H_2O/CH_4$ absorption feature at 1.4 μm and the $H_2O$ absorption feature at 1.8 μm are much harder to resolve even though the red Howe & Burrows model makes it seem otherwise. Extrapolating this case out to the other masses and temperatures, it can be deduced that with 25 transits, an H-rich planet could be observed at M ≥ 1 M⊕ and T ≥ 400 K out to 11 pc. For planets with larger masses and higher temperatures it would be possible to observe a H-rich planet beyond 50 pc if it had a M ≥ 10 M⊕ and/or T ≥ 1000 K.

The H-poor case presents very different results. Looking at the case of the 10 M⊕, 1000 K, H-poor planet in the upper right hand panel of **Figure 3**, there is still clear detection of $H_2O$ and $CH_4$ at all the wavelengths mentioned above at 4.5 pc (black curve). It is not possible to observe this planet out to 50 pc, as with the H-rich case. It is only possible to obtain a SNR of 15 with 25 transits at 26 pc. The boxes shaded in red are cases where an SNR of 15 was not achievable. Therefore, the spectrum of any H-poor planet with T ≤ 400 K would be dominated by noise. Only if the H-poor planet has T > 700 K, will it be possible to resolve any of the water bands especially at M ≤ 4 M⊕.

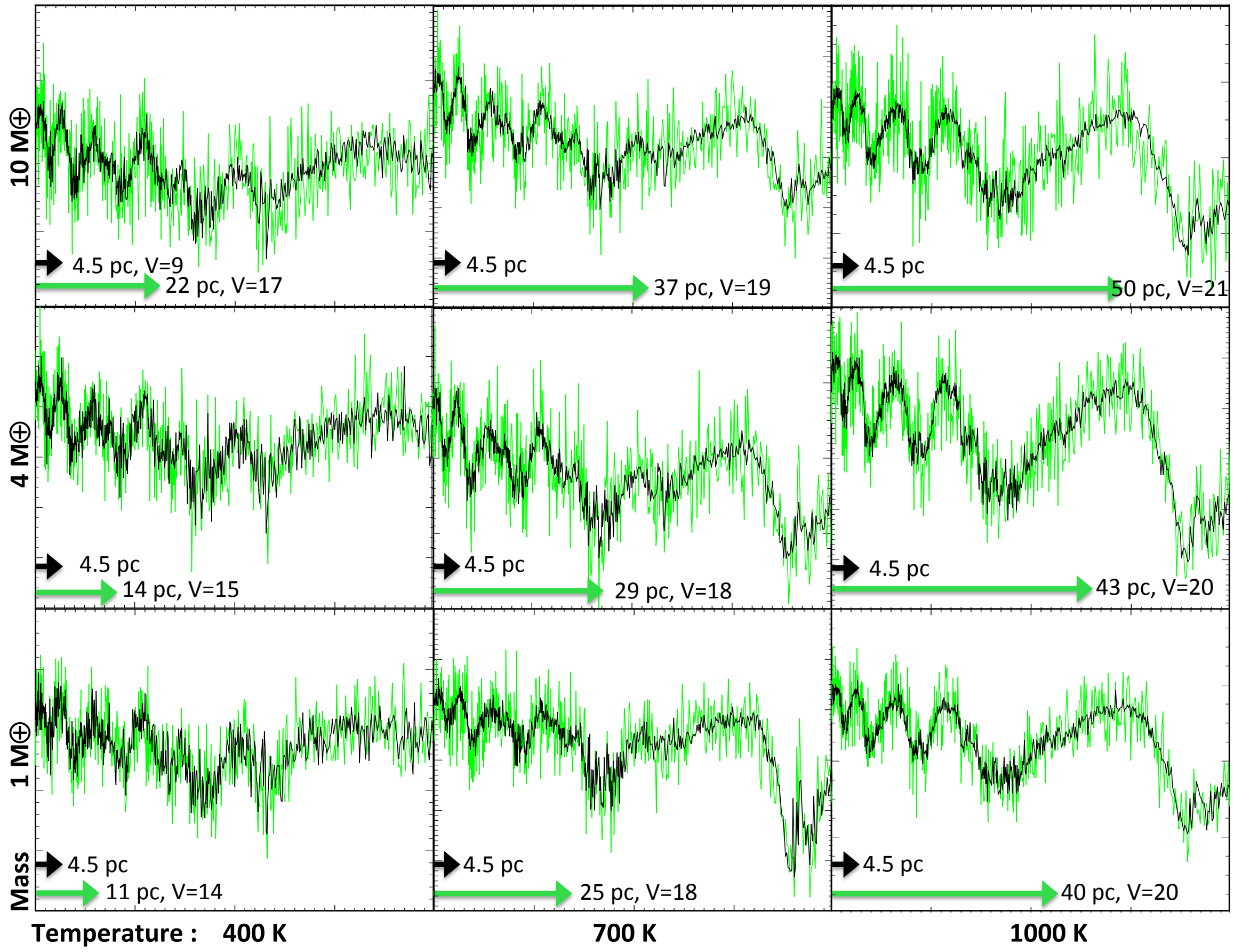

**Figure 2. Hydrogen Rich Atmosphere-** Simulated JWST/NIRSpec spectra at three different temperatures (400, 700, and 100 K) and three different masses (1, 4, and 10 M⊕). All simulations assume 25 transits and each panel span 1 – 5 μm in wavelength. The simulations include effects of pointing jitter, flat field errors and pixel response function. Model spectra were taken from Howe & Burrows (2012) for 3x solar metallicity, assuming a parent star with similar properties to GJ1214 (V=14.67) except with varying distances from 4.5 – 50 pc. The spectra in black illustrate the most optimistic case at 4.5 pc, whereas the green curve shows the furthest distance you can achieve SNR ~ 15. For this case of a H-rich atmosphere, JWST/NIRSpec can provide excellent characterization of water and methane absorption features at 2.7 and 4.6 μm out to 11 pc at $M \geq 1$ M⊕ and $T \geq 400$ K.

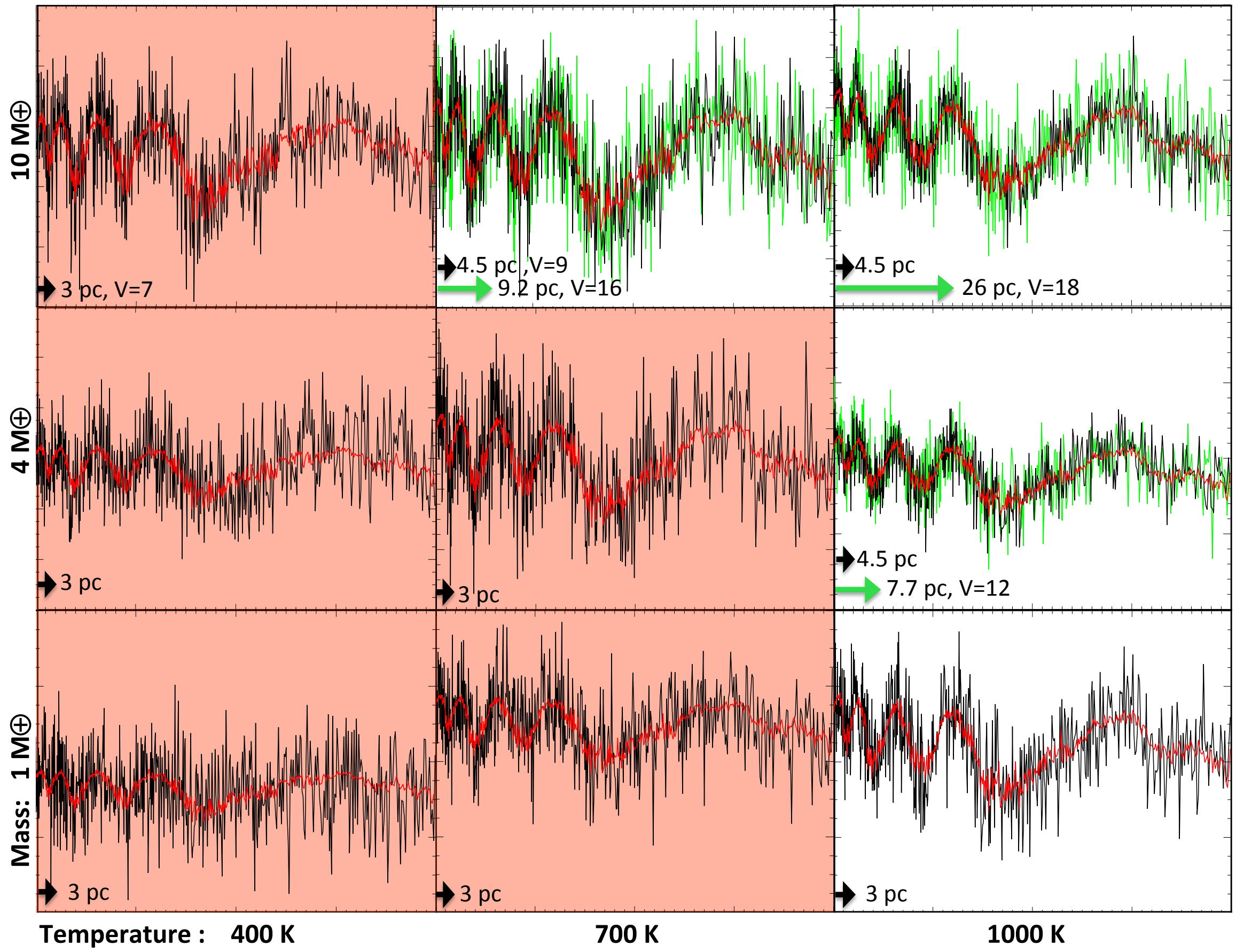

**Figure 3. Hydrogen Poor Atmosphere-** Simulated JWST/NIRSpec spectra at three different temperatures (400, 700, and 100 K) and three different masses (1, 4, and 10 M⊕). All simulations assume 25 transits and each panel span 1 – 5 μm in wavelength. The simulations include effects of pointing jitter, flat field errors and pixel response function. Model spectra (red) were taken from Howe & Burrows (2012) for pure water atmospheres, assuming a parent star with similar properties to GJ1214 (V=14.67) except with varying distances from 3 – 26 pc. The spectra in black illustrate the most optimistic case at 4.5 or 3 pc, whereas the green curve shows the furthest distance you can achieve SNR ~ 15. For this case of a H-poor atmosphere, JWST/NIRSpec will be limited to high-signal to noise characterization of more massive and/or hotter super-Earths (yellow and white boxes). The red shaded boxes are cases where it is not possible to achieve SNR of 15, even at 3 pc.

## 5. Summary and Future Work

JWST will be astronomy's most space powerful telescope to date, and it is poised to transform our understanding of the Universe. In this white paper, we provided realistic simulations of JWST to characterize the increasing population of nearby super-Earth exoplanets orbiting M dwarfs.

Our results are based on simulations derived from a JWST simulator combined with the latest exoplanet transit models from Howe & Burrows (2012). We executed a small sample of simulations of primary transits, studied the changes in SNR, and interpolated these results over a larger parameter space (mass, temperature, and distance). Based on a detectability criterion of an SNR of ~15, we determine that 25 transits on JWST/NIRSpec is sufficient to detect water and methane bands at 2.7 and 4.6 μm out to 11 pc at $M \geq 1 M_{\oplus}$ and $T \geq 400$ K. For the H-poor case at 25 transit observations, a temperature above 700 K will be necessary to resolve any water bands, especially for planets with $M < 4 M_{\oplus}$.

Following the promising results from this project, we will expand our study to repeat these calculations with the NIRISS instrument on JWST, to expand the set of host stars to G and K spectral types, and to include additional atmosphere models (e.g., intermediate between H-poor and H-rich). Modeling of MIRI spectra from primary transits, and examination of secondary transit spectra, constitute a separate future effort.

We thank Avi Mandell and Mark McCaughrean for helpful conversations, comments, and suggestions.

**References**

Anglada-Escudé, G., Arriagada, P., Vogt, S. S., et al. 2012, ApJ, 751, L16
Batalha, Natalie M., Kepler Team, 2013, ApJ, 204, 24
Charbonneau, D., Berta, K., Irwin, J., et al. 2009, Nature, 462, 891
Clampin, M. Comparative Planetology, JWST Whitepaper 2010
Crouzet, N., McCullouhj, P. R., Burke, C., Long, D. 2012, ApJ, 761, 7
Deming, D., Harrington, J., Lauglin, G., 2007, ApJ, 667, L199
García-Muñoz, A., Zapatero Osorio, M.R., Barrena, R., Montañés-Rodríguez, P., Martín, E.L. and Pallé, E.2012. ApJ. 755, 103.
Kaltenegger, L., Udry, S., Pepe, F., 2011, eprint arXiv: 1108.3561
Kopparapu, R., Ramirez, R., Kasting, J., et al. 2013, ApJ, 765, 131
Mayor, M., Udry, S., Lobis, C., et al. 2009, A&A, 493, 639
Miller-Ricci, E., Seager, S., Sasselov, D., 2009, ApJ, 690, 1056
Pepe, F., Lovis, C., Ségransan, D., et al. 2011, A&A, 534, A58
Ricker, G. R., Latham, D. W., Vanderspek, R. K., et al. 2010 AAS 21545006
Sing, D. K., Point, F., Aigrain, S., et al. 2011, MNRAS, 416, 1443
Tuomi, M., Jones, H. R. A., Jenkins, J. S., et al. 2013, A&A, 551, A79
Vogt, S. S., Butler, R. P., Rivera, E. J., et al. 2010, ApJ, 723, 954